\documentclass{desyproc}

\begin{document}
\title{Photons and Exclusive Processes at Hadron Colliders}

\author{{\slshape Joakim Nystrand}\\[1ex]
Department of Physics and Technology, University of Bergen, Bergen, Norway }

\contribID{47}

\confID{1407}  
\desyproc{DESY-PROC-2009-03}
\acronym{PHOTON09} 
\doi  

\maketitle

\begin{abstract}
The theoretical and experimental aspects of particle production from the
strong equivalent photon fluxes present at high energy hadron colliders
are reviewed. The goal is to show how photons at hadron colliders
can improve what we have learnt from experiments with lepton beams.
Experiments during the last 5-10 years have shown the feasibility of
studying photoproduction in proton-proton and heavy-ion collisions. The
experimental and theoretical development has revealed new opportunities as
well as challenges.
\end{abstract}

\section{Introduction}

During the last decade the idea to study electromagnetic particle production at hadron and 
heavy-ion colliders has developed theoretically as well as experimentally. At the previous 
conference in this series (Photon 07, held at the Sorbonne in Paris), a session was devoted 
to this topic. 
I gave a summary of photoproducion at hadron colliders, in particular in heavy-ion collisions, 
at that meeting\cite{Nystrand:2008qw}. Since then, several new results have been published 
and an update of the latest development will be given here. More comprehensive reviews of the 
field can be found e.g. in \cite{Bertulani:2005ru} and \cite{Baltz:2007kq}.

There are several reasons to study photon-induced reactions in collisions between hadrons. At the 
Large Hadron Collider (LHC) at CERN, the accessible two-photon center of mass energies 
will be much higher than at LEP and the photon-nucleon energies will be higher than at 
HERA. The LHC will thus explore a new and uncharted territory also in interactions 
mediated by photons. Heavy-ion interactions at the Relativistic Heavy-Ion Collider 
and at the LHC provide an opportunity to study photons from strong electromagnetic fields 
where the coupling is $Z \sqrt{\alpha}$ rather than $\sqrt{\alpha}$. At a
hadron collider both projectiles can act as photon emitter and target, and this leads to some 
interesting interference phenomena, which will be discussed below. Furthermore, the 
electromagnetic contribution is important for understanding other types of exclusive 
particle production, e.g. through Pomeron+Pomeron interactions in $p p$ 
or $p \overline{p}$ collisions. It has been been proposed to use two-photon 
production of $\mu^+ \mu^-$--pairs as a luminosity monitor at the LHC. 

To separate the electromagnetic interactions from the ``background'' of strong, hadronic 
processes, it is in practice necessary to restrict the study to so-called ultra-peripheral 
collisions, where the impact parameters are larger than the sum of the beam particle radii. 
This means impact parameters in the range $\approx$1.4(14) - 100~fm in proton-proton (heavy-ion 
collisions). The upper range 100 fm is somewhat arbitrary but in practice the fields at current 
colliders are too weak to produce particles outside this range. The exception is production of 
low-mass $e^+ e^-$--pairs, where impact parameters above 1000~fm can give a significant contribution. 
Experimentally, interactions mediated by photons can be separated from purely hadronic interactions  
by their lower multiplicity and the presence of rapidity gaps. If the entire event is 
reconstructed, charge conservation and the low total transverse momentum also provide strong 
background rejection.

\section{Results from RHIC}

Ultra-peripheral collisions have been studied at the Relativistic Heavy-Ion Collider 
(RHIC) at Brookhaven National Laboratory by the two large experiments PHENIX and STAR. 
The first results on $\rho^0$ photoproduction were published by the STAR collaboration 
based on data from the first run at RHIC in the year 2000. STAR has so far mainly focussed  
on low-mass states, such as photoproduction of the $\rho^0$ meson and low-mass 
$e^+e^-$--pairs. The study of ultra-peripheral collisions in PHENIX has been directed 
towards heavy vector meson production (J/$\Psi$) and high-mass $e^+e^-$--pairs. The first
preliminary results from PHENIX were presented at the Quark Matter 2005 
conference\cite{d'Enterria:2006ep}, 
and earlier this year the final results were published\cite{Afanasiev:2009hy}. 
The final results also include cross sections for two-photon production of 
$e^+ e^-$--pairs. 

\begin{figure}[hb]
\centerline{\includegraphics[width=0.45\textwidth]{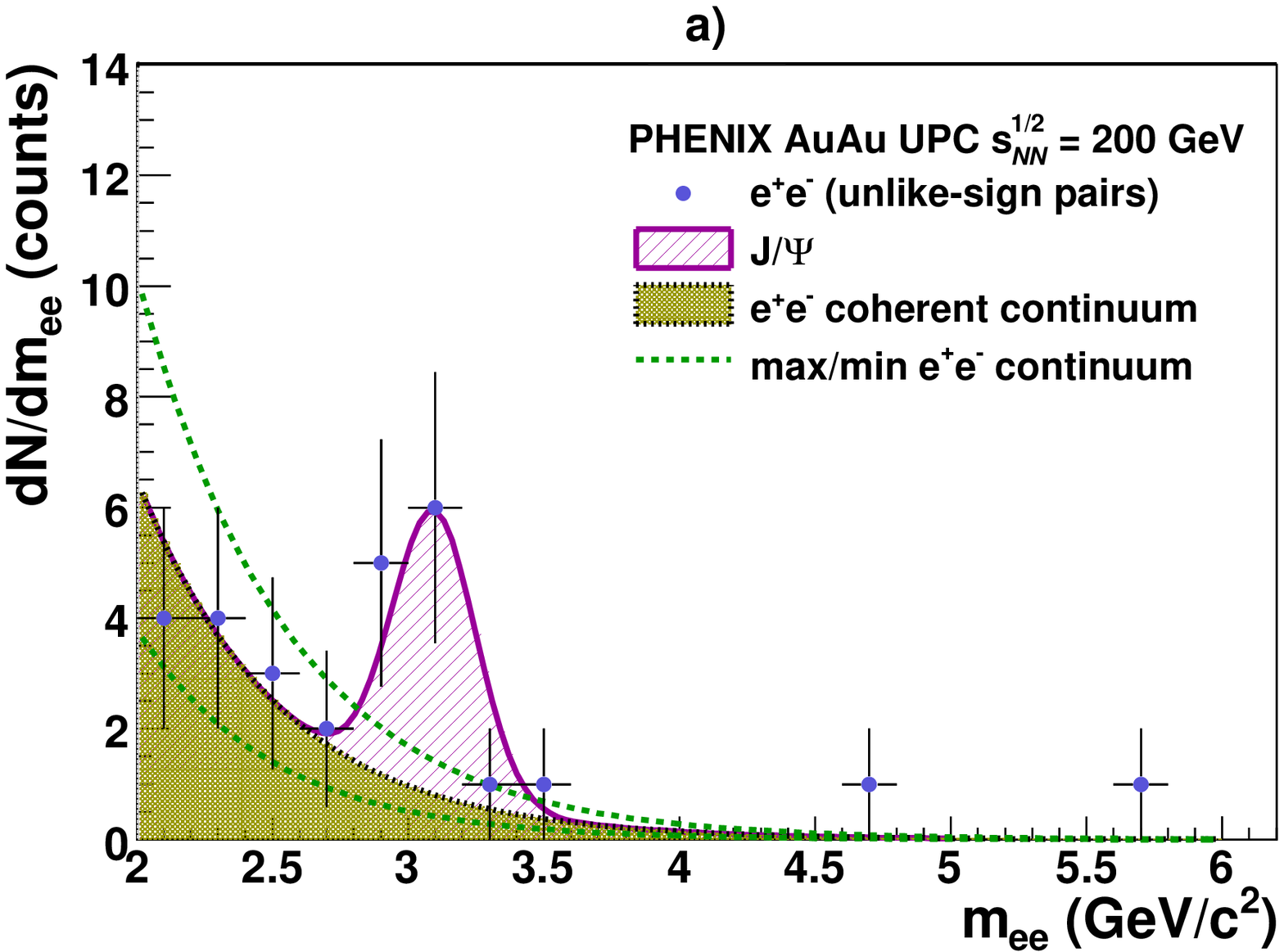}
\includegraphics[width=0.45\textwidth]{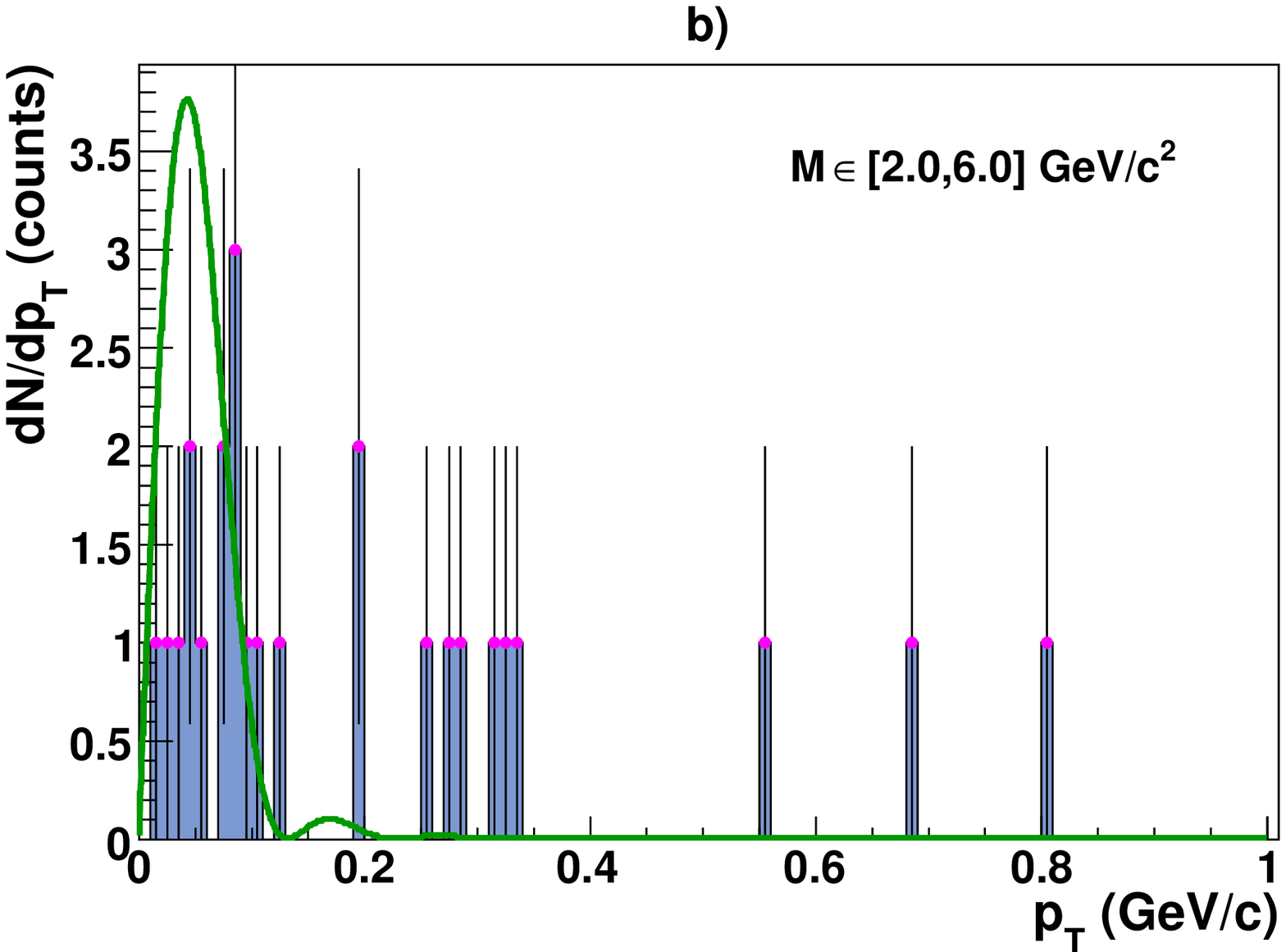}}
\caption{Invariant mass (a) and transverse momentum (b) distributions of $e^+ e^-$--pairs in ultra-peripheral 
Au+Au collisions as measured by PHENIX\cite{Afanasiev:2009hy}. 
The event selection is described in the text.}\label{Fig:phenix}
\end{figure}

PHENIX has implemented a trigger where an energy deposit of $E > \approx 0.8$ GeV in the 
mid-rapidity electromagnetic calorimeters is combined with two rapidity gaps at intermediate 
rapidities. 
The rapidity gaps are defined by an absence of a signal in the Cherenkov Beam-Beam Counters (BBC) 
covering the pseudorapidity range $3.0 < |\eta| < 3.9$ on either side of midrapidity. 
It is furthermore required 
that there should be a signal ($E > \approx 30$~GeV) in one or both of the Zero-Degree Calorimeters 
(ZDC) situated 18 m 
downstream from the interaction point. This requirement means that the central production 
occurs in coincidence with exchange of one or more additional photons which lead to the break up 
of one or both nuclei. 

A data sample with an integrated luminosity of $141\pm12 \mu$b from the 2004 high-luminosity 
run at RHIC has been analyzed. This luminosity corresponds to 
around $960 \cdot 10^6$ inelastic Au+Au events ($\sigma_{inel} = 6.8$~b). The UPC trigger as described 
above collected $8.5 \cdot 10^6$ events out of which $6.7 \cdot 10^6$ satisfied standard data 
quality assurance criteria. The electrons and positrons were tracked and identified in the PHENIX 
central tracking arms, covering $|\eta| <0.35$ and $\Delta \phi = 90^o$ each. Tracking was 
performed by multi-layer drift chambers and multi-wire proportional chambers in each arm, and  
electrons were identified by their signals in the Ring-Imaging-Cerenkov (RICH) detectors and the  
electromagnetic calorimeters. 
In the offline analysis, it was required that the event should contain exactly 2 tracks 
of opposite charge satisfying the UPC electron/positron identification criteria and that 
the event vertex should be within $|Z| < 30$~cm from the center of the detector. 
It was required that each electron/positron should have 
$E > 1$~GeV to be well above the trigger threshold ($E > \approx 0.8$~GeV). 
This gave a sample with 28 events, the invariant
mass and $p_T$ distributions of which are shown in Fig.~\ref{Fig:phenix}. There was no like-sign 
background after the offline cuts had been applied. The results are consistent 
with production of $e^+ e^-$--pairs through two-photon interactions, 
$\gamma+\gamma \rightarrow e^+ e^-$, and photonuclear production (coherent and incoherent) of $J/\Psi$, 
$\gamma + Au \rightarrow J/\Psi + Au$, followed by the decay $J/\Psi \rightarrow e^+ e^-$. The 
background from hadronically produced $e^+ e^-$--pairs in the ultra-peripheral sample 
was estimated from measurements of $e^+ e^-$--pairs in proton-proton collisions and 
was found to be negligible ($< 1$~event). The majority of the events are from coherent production,
as can be seen from the transverse momentum distribution in Fig.~\ref{Fig:phenix} b). The curve 
shows for comparison the nuclear (Au) form factor. 

The cross section for $J/\Psi$ production, $d\sigma/dy = 76 \pm 31 (stat) \pm 15 (syst) \mu$b 
with nuclear break up, is consistent with theoretical 
predictions\cite{Klein:1999qj,Strikman:2005ze,Ivanov:2007ms,Goncalves:2005sn}, but the 
size of the experimental errors preclude a discrimination between the models.  
The cross section for continuum $e^+ e^-$ production are shown in 
Table~\ref{tab:phenix}. The results are in good agreement with calculations using the method of 
equivalent photons as implemented in the Starlight Monte Carlo\cite{Baltz:2009jk}. 

The mid-rapidity cross section for $J/\Psi$ is a measure of the $J/\Psi +$~Nucleon cross section 
in nuclear matter\cite{Strikman:2005ze} and is also affected by nuclear gluon 
shadowing\cite{AyalaFilho:2008zr}. Recent calculations by Baltz\cite{Baltz:2007gs} for two-photon
production of di-muons and di-electrons have found that higher order terms reduce the 
cross section by 20-30\% in heavy-ion collisions. 
A recent calculation has however not found such large reductions from higher order terms, 
particularly not for $\mu^+ \mu^-$--pairs\cite{Jentschura:2009mb}. A measurement with smaller 
errors could settle this issue experimentally. 

\begin{table} 
\begin{center} 
\begin{tabular}{|ccc|} \hline
$m_{inv}$ $[GeV]$ & $d\sigma/dm_{inv}dy$ & $d\sigma/dm_{inv}dy$ \\ 
                  & $(|y|<0.35)$         & $(|y|<0.35, |\eta_{1,2}|<0.35)$ \\ \hline 
$2.0 \leq m_{inv} \leq 2.8$ & $86 \pm 23 (stat) \pm 16 (syst)$ & $0.95 \pm 0.25 (stat) \pm 0.18 (syst)$  \\ 
$2.0 \leq m_{inv} \leq 2.3$ & $129 \pm 47 (stat) \pm 28 (syst)$ & $1.43 \pm 0.52 (stat) \pm 0.31 (syst)$ \\  
$2.3 \leq m_{inv} \leq 2.8$ & $60 \pm 24 (stat) \pm 14 (syst)$ & $0.65 \pm 0.26 (stat) \pm 0.15 (syst)$ \\ \hline  
\end{tabular}
\caption{Differential cross sections for two-photon production of $e^+ e^-$--pairs\cite{Afanasiev:2009hy}. 
The variables $y$ and $m_{inv}$ refer, respectively, to the rapidity and invariant mass of the pair. 
The rightmost column shows the cross section when one requires both the $e^+$ and $e^-$ to be within 
$|\eta| <$~0.35.}
\label{tab:phenix}
\end{center}
\end{table}

\begin{figure}[bh]
\centerline{\includegraphics[width=0.45\textwidth]{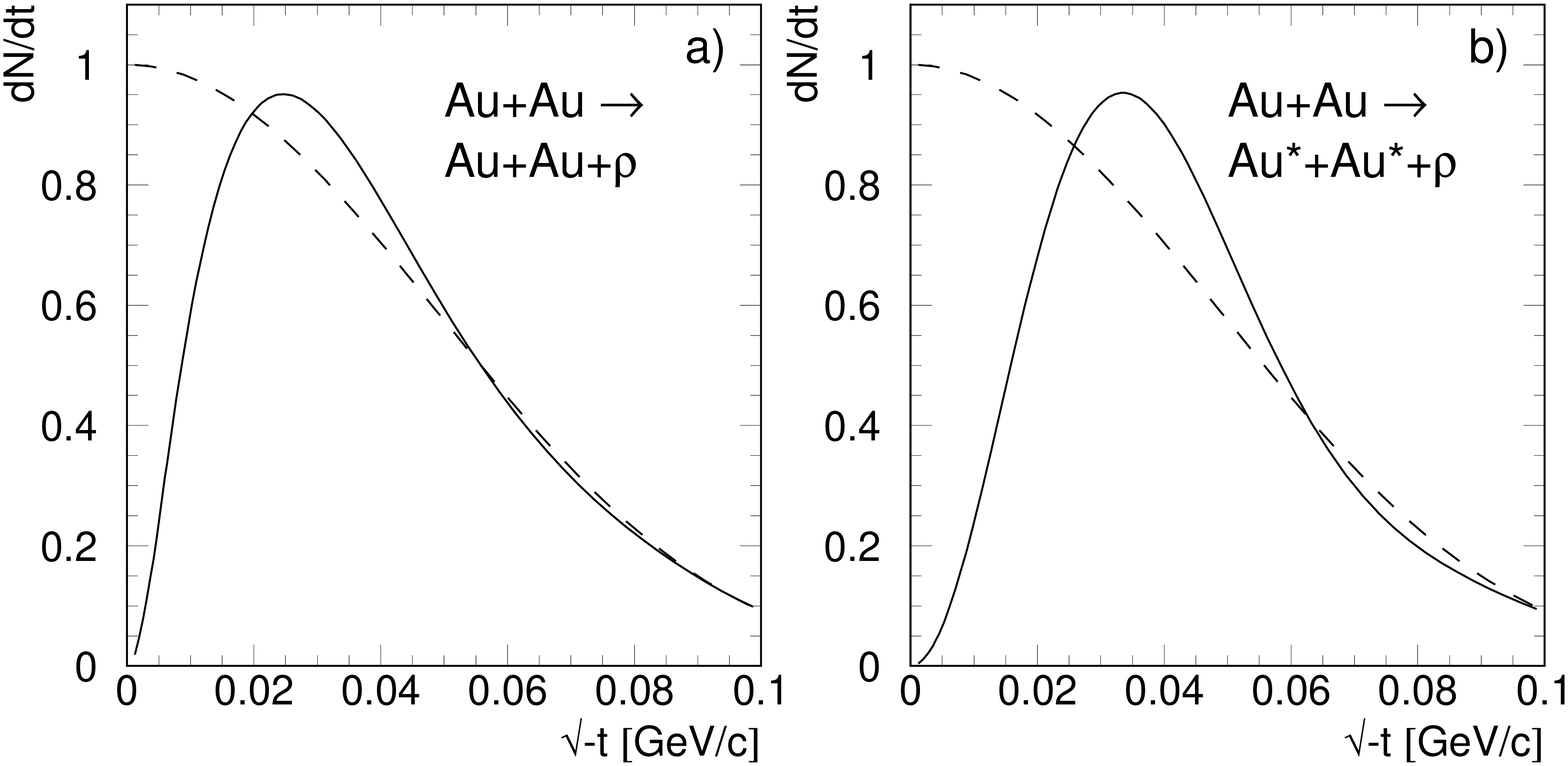}}
\caption{Predicted $\rho^0$ transverse momentum spectra with (solid curve) and without (dashed curve) 
interference\cite{Klein:1999qj,Klein:1999gv}. The left plot shows the spectrum for all events 
while the right plot shows the spectrum for events with mutual Coulomb break up.}
\label{Fig:interference}
\end{figure}

The STAR collaboration has studied coherent and incoherent photoproduction 
of $\rho^0$ mesons in Au+Au collisions at $\sqrt{s_{NN}} =$ 130 and 
200 GeV\cite{Adler:2002sc,Abelev:2007nb}. 
The pions from the decay $\rho^0 \rightarrow \pi^+ \pi^-$ were reconstructed in the 
STAR Time-Projection Chamber (TPC), 
which covers the pseudorapidity range $| \eta | <$~1 with full coverage in azimuth. 
Two trigger classes were defined. One class was based on having two 
roughly back-to-back hits in the Central Trigger Barrel (CTB), a set of 240 scintillators 
surrounding the TPC. This trigger (``topology trigger'') selects photoproduced $\rho^0$ 
mesons with and without nuclear break up. 
The second trigger class (``minimum bias (MB)'') required the events to have a signal 
in both ZDCs, and thus selected photoproduction in coincidence with mutual 
Coulomb breakup of both Au-nuclei. 

The cross sections, rapidity and $p_T$ distributions have been published earlier and 
have been found to be in 
good agreement with calculations\cite{Adler:2002sc,Abelev:2007nb}.  

\begin{figure}[ht]
\centerline{\includegraphics[width=0.45\textwidth]{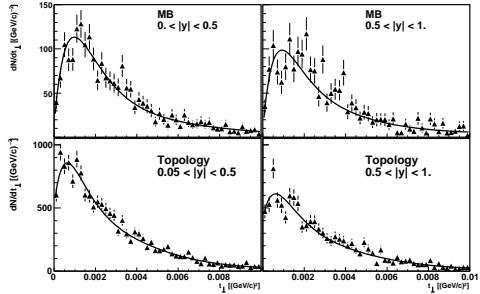}}
\caption{Transverse momentum ($p_T^2 = t_{\perp}$) spectrum for photoproduction 
of $\rho^0$ mesons in ultra-peripheral Au+Au collisions as measured by STAR\cite{Abelev:2008ew}.
The top plots are for minimum bias data and the bottom plots are for the topology trigger.
The plots in the left column show production at mid-rapidity and those in the right column 
show production away from mid-rapidity. The curves are fits to a function with an  
interference term.}\label{Fig:star}
\end{figure}

Both nuclei can act as photon emitter and target in an ultra-peripheral collision.  
Since the two possibilities cannot be distinguished, they might interfer quantum mechanically. 
The separation of the two nuclei when a $\rho^0$ meson is produced at RHIC energies is typically 
14--40 fm, with a different distribution depending on if the $\rho^0$ is produced with or 
without nuclear breakup. The median impact parameters are $<b> =$~46~fm for the total production 
and $<b> =$~18~fm with mutual break up\cite{Klein:1999qj}. The $\rho^0$ meson is always 
produced near the surface of one of the nuclei because of the short range of the nuclear force. 
The system thus acts as a two-source interferometer and, depending on the separation of the nuclei 
and the wave length of the $\rho^0$, interference might occur. The interference will be 
destructive in collisions between particles and constructive in collisions between 
particles and anti-particles. For production at mid-rapidity, where the 
amplitudes for the two possibilities are equal, the cross section for a meson with tranverse 
momentum $\vec{p_T}$ in a collision with impact parameter $\vec{b}$ is 
\begin{equation}
\sigma(\vec{p_T},\vec{b}) = 2 \sigma_1(\vec{p_T},\vec{b}) \left( 1 \pm cos(\vec{p_T} \cdot \vec{b}) \right) \; ,
\end{equation}
where $\sigma_1(\vec{p_T},\vec{b})$ is the cross section for emission from a single source. 
The total production cross section is obtained by integrating over all allowed impact parameters. 
For $p_T << 1/<b>$, the cross section will be affected by the interference. This is illustrated in 
Fig.~\ref{Fig:interference}, where the solid curve shows the expected $t = p_T^2$ spectrum with 
interference and the dashed curve shows the expected spectrum without interference. The smaller 
mean impact parameters in collisions with nuclear break means that the interference extends 
to higher transverse momenta, as seen in Fig.~\ref{Fig:interference} b). 

The presence of interference and its effect on the vector meson transverse momentum spectrum   
was predicted\cite{Klein:1999gv}. It has now been experimentally confirmed by the STAR 
collaboration\cite{Abelev:2008ew}.
The measured transverse momentum distributions for the two trigger classes are shown in 
Fig.~\ref{Fig:star}. As one moves away from mid-rapidity, the amplitudes for the two 
target and photon-emitter configurations will be different and one expects the interference 
to decrease. This can be seen in the two plots to the right in Fig.~\ref{Fig:star}. 

What makes this interference particularly interesting is the fact that the life-time 
of the $\rho^0$ is much smaller than the separation between the nuclei divided by the 
speed of light. The interference must therefore occur between the decay products of the 
$\rho^0$, which are in an entangled state. The interference is the result of the large 
separation between the nuclei in an ultra-peripheral collision and is expected to be 
different for other coherent production mechanisms, e.g. Odderon--Pomeron fusion.

\section{Results from the Tevatron} 

Two-photon and photonuclear interactions have been studied by the CDF collaboration 
in proton-anti-proton collisions at the Tevatron. The latest results include 
exclusive di-muon production in the invariant mass 
range 3~$\leq m_{\mu \mu} \leq$~4 GeV\cite{Aaltonen:2009kg}  
and two-photon production of di-leptons with 
$m_{l l} >$~40 GeV\cite{Aaltonen:2009cj}. The produced particles have been detected at 
central rapidities, $|\eta| <$~0.6 for intermediate mass di-muons and $|\eta| <$~4 for the 
high mass di-leptons. The centrally produced particles are observed in an otherwise empty 
event, which is defined by other detectors with a coverage out to $|\eta| <$~7.4. Further 
details on the experimental setup, trigger, and the exact phase space coverage can be found 
in Refs. \cite{Aaltonen:2009kg,Aaltonen:2009cj}. 

The exclusive di-muon events were identified as coming from exclusive photoproduction of 
$J/\Psi$ and $\Psi'$ followed by the decay to $\mu^+ \mu^-$ and two-photon 
continuum production of $\mu^+ \mu^-$--pairs. The background from the $\chi_c$ mesons 
produced in exclusive Pomeron-Pomeron interactions decaying to $\chi_c \rightarrow J/\Psi + \gamma$ 
was estimated. Measurements were made where a $J/\Psi$ was produced 
in coincidence with a photon with energy $E_{\gamma} >$~80 MeV. From this, it could be estimated 
that the background from $\chi_c$ events where the photon is not detected contributed about 4\%
to the exclusive $J/\Psi$ sample. 

\begin{figure}[hb]
\centerline{\includegraphics[width=0.65\textwidth]{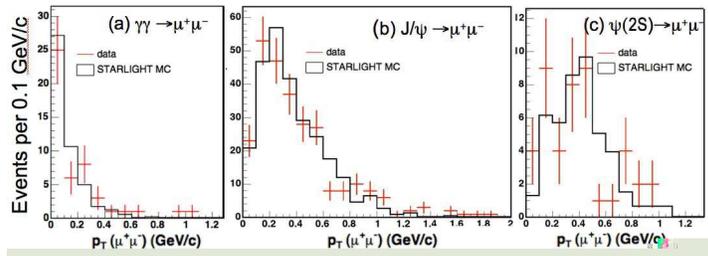}}
\caption{Transverse momentum distributions of exclusive $\mu^+ \mu^-$--pairs in $p \overline{p}$ 
collisions, as measured by the CDF collaboration\cite{Aaltonen:2009kg}.
The crosses show the experimental results and the histograms are the Starlight predictions, 
normalized to the data.
Fig. a) shows $\mu^+ \mu^-$--pairs in the invariant mass ranges $3.2 \leq m_{inv} \leq 3.6$ 
and $3.8 \leq m_{inv} \leq 4.0$~GeV.}\label{Fig:cdf}
\end{figure}

The exclusive events can be simulated with the Starlight Monte Carlo, which is based on the 
model in\cite{Klein:2003vd}. The photon-proton cross section as measured by experiments at HERA 
and at fixed target experiments with lepton beams are used as input to the calculations. These 
are combined with the equivalent photon spectrum to give the cross section for 
$p+\overline{p} \rightarrow p+\overline{p}+V$, where V is a vector meson. Starlight can also calculate the 
cross section for $\mu^+ \mu^-$ pairs produced in two-photon interactions. 
The differential cross section, $d\sigma/dy$, for vector meson production is given by 
\begin{equation}
\begin{array}{lcr}
\frac{d \sigma(p + \overline{p} \rightarrow p + \overline{p} + V)}{dy} & = & \\
k_1 \frac{dn_{\gamma}}{dk_1} \sigma_{\gamma p } (k_1) & + &
k_2 \frac{dn_{\gamma}}{dk_2} \sigma_{\gamma p } (k_2) . \\
\end{array}
\end{equation}
Here, $dn/dk$ is the photon spectrum and $\sigma_{\gamma p }$ the cross section for 
$\gamma + p \rightarrow V + p$. 
A $J/\Psi$ produced at central rapidity within $|y| <$~0.5 corresponds to photoproduction with 
photon-proton center of mass energies between $60 \leq W_{\gamma p} \leq 100$~GeV. Exclusive 
$J/\Psi$ and $\Psi'$ production has been studied at HERA in this energy range by both 
the Zeus\cite{Chekanov:2002xi} and H1\cite{Aktas:2005xu} collaborations. The 
$\gamma + p \rightarrow J/\Psi + p$ cross sections are thus know experimentally, with errors 
typically in the range 6--9 \%. 

The photon spectrum associated with a relativistic proton can be calculated from that of a 
point charge modulated by a form factor\cite{Drees:1988pp}. 
A problem with this approach is that it does not properly exclude collisions where the protons 
interact hadronically. A different approach is therefore used here. 
The method is similar to that used for nuclear collisions, where one has to 
require that the impact parameter be larger than the sum of the nuclear radii to exclude 
strong interactions\cite{Cahn:1990jk}. 
The proton has a more diffuse surface, however, so applying a sharp cut-off 
in impact parameter space is unphysical. A more realistic approach is to calculate the 
(hadronic) interaction probability as function of impact parameter by applying a Fourier transform 
to the $pp$ elastic scattering amplitude\cite{Frankfurt:2006jp}. To set a conservative 
upper limit on the photon spectrum, it is also calculated with a cut on impact parameter $b >$~0.7 fm. 
This gives a photon spectrum very similar to that obtained when using the proton form factor. 

The transverse momentum distribution for the three final states are shown in Fig.~\ref{Fig:cdf} 
together with predictions by Starlight. The transverse momenta of the vector mesons reflect 
the proton form factor and extend out to around $\approx 1$~GeV/c, whereas the transverse momenta 
of the two-photon final state are considerably lower. The agreement between data and Starlight is 
very good.  

For the calculations of the cross sections, the latest results from HERA were used to make a 
new fit to the $\gamma + p \rightarrow V+p$ cross sections. These fits are slightly different from 
the ones used earlier\cite{Klein:2003vd}. The uncertainty in the calculated cross sections are 
determined from the uncertainty in the measured $\gamma p$ cross sections and the calculation 
of the photon spectrum with $b >$~0.7 fm. The results are 
\begin{equation}
J/\Psi: \;\; \left. \frac{d\sigma}{dy} \right|_{y=0} = 2.7^{+0.6}_{-0.2} \; nb \;\;\;  \;\;\; \;\;\; , \;\;\; \;\;\; \;\;\;
\Psi': \;\;  \left. \frac{d\sigma}{dy} \right|_{y=0} = 0.45^{+0.11}_{-0.04} \; nb \; .
\end{equation}
These can be compared with the measured values, $3.92 \pm 0.25 (stat) \pm 0.52 (syst)$~nb and 
$0.53 \pm 0.09 (stat) \pm 0.01 (syst)$~nb, for the $J/\Psi$ and $\Psi'$, 
respectively\cite{Aaltonen:2009kg}. The measured cross section for the $J/\Psi$ is thus 
about two standard deviations above the calculated value and about one standard deviation 
above the calculated upper limit. The CDF collaboration has concluded that the upper limit 
for an Odderon contribution (Odderon+Pomeron$\rightarrow J/\Psi$) is less than 
$d\sigma(y=0)/dy <$~2.3 nb with 95\% confidence level.

\section{Outlook to the LHC}

The results from RHIC and the Tevatron shows the feasibility of studying two-photon and 
photon-nucleon interactions at hadron colliders. The results have been found to be in 
general agreement with predictions, but the statistics at least for heavy final states 
has so far been rather low. 

\begin{figure}[bh]
\centerline{\includegraphics[width=0.55\textwidth]{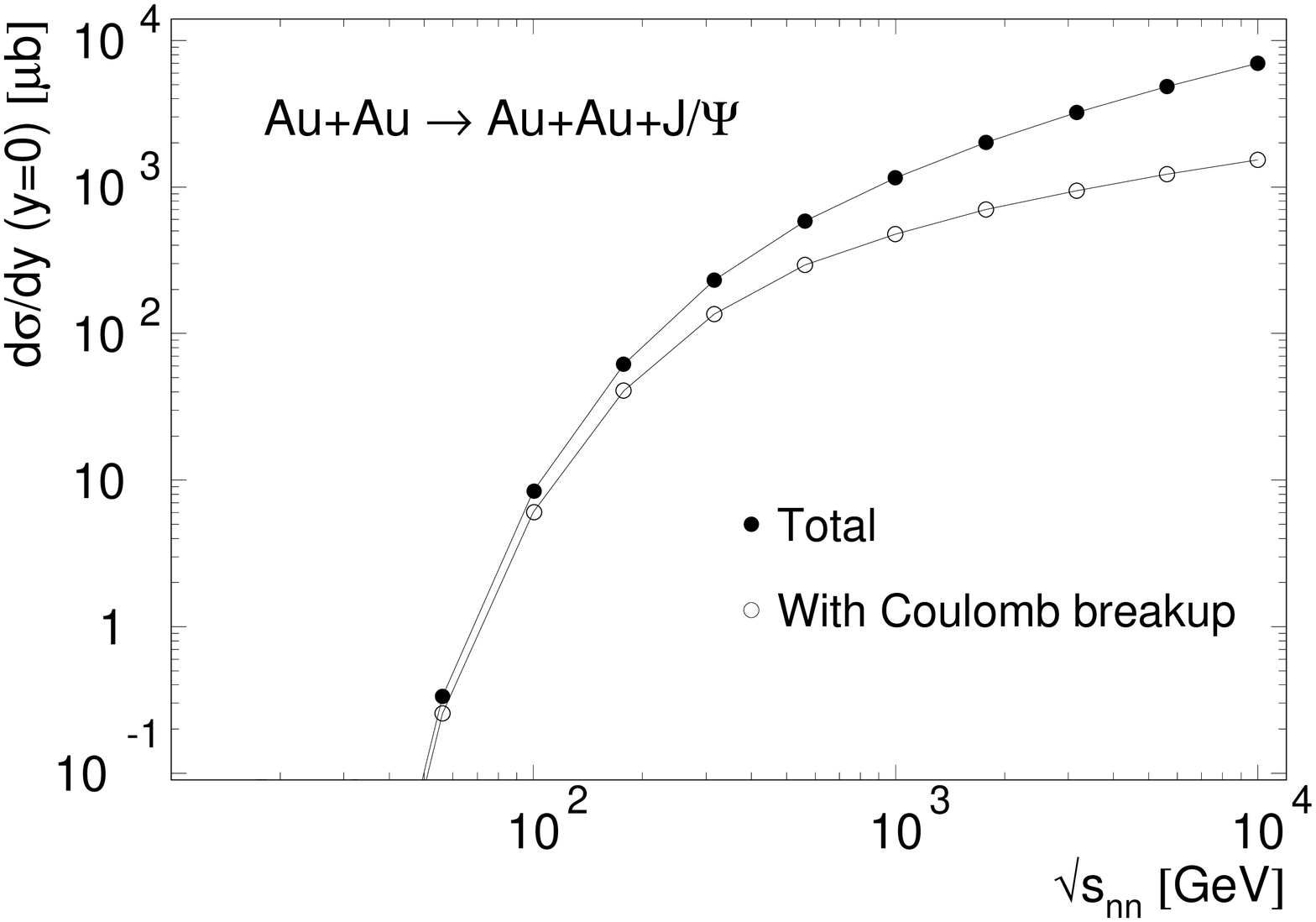}}
\caption{Calculated excitation function for mid-rapidity photoproduction of $J/\Psi$ in Au+Au 
collisions, based on \cite{Klein:1999qj}.}\label{Fig:lhc}
\end{figure}

The situation should be more advantageous at the LHC for at least two reasons: First, the 
cross section increases dramatically with the increased collision energy. This is 
illustrated in Fig.~\ref{Fig:lhc}, which shows the excitation function for $J/\Psi$ 
production at mid-rapidity in heavy-ion collisions. The increase in cross section is 
about a factor of 100 between RHIC and LHC energies. Secondly, the large and versatile 
experiments at the LHC should have the capability to trigger on and reconstruct 
particles produced in ultra-peripheral collisions over a wide range of phase space. 

Several topics can be studied in ultra-peripheral collisions at the LHC. 
The CMS collaboration has for example investigated the possibilities for studying 
exclusive $\Upsilon$ production in Pb+Pb collisions\cite{D'Enterria:2007xr} and 
two-photon production of supersymmetric pairs in pp collisions\cite{Schul:2008sr}. 
ALICE, which is primarily aimed for studying heavy-ion collisions, has performed 
simulations of $J/\Psi$ and two-photon production in heavy-ion and proton-proton 
collisions\cite{Alessandro:2006yt}. 

Most studies so far, both at the existing colliders RHIC and the Tevatron as well 
as at the LHC, have focussed on exclusive production of a single particle or pair 
of particles. It should however be possible to extract as much information from 
inclusive processes, for example heavy quark production from photon-gluon fusion 
and photon-induced jet production. 

To summarize, ultra-peripheral collisions are an interesting enhancement of the physics 
programs at hadron colliders. They can be studied at existing and planned experiments 
with no or only minor modifications. The results from RHIC and the Tevatron have 
followed the theoretical expectations. Further studies at the LHC and at existing accelerators 
with increased luminosities will hopefully lead to useful constraints on e.g. the 
parton density distributions and possibly even the discovery of new phenomena.



\begin{footnotesize}




\begin{thebibliography}{99}

\bibitem{Nystrand:2008qw}
J.~Nystrand,
Nucl.\ Phys.\ Proc.\ Suppl.\  {\bf 184} (2008) 146.

\bibitem{Bertulani:2005ru}
C.~A.~Bertulani, S.~R.~Klein and J.~Nystrand,
Ann.\ Rev.\ Nucl.\ Part.\ Sci.\  {\bf 55} (2005) 271.

\bibitem{Baltz:2007kq}
K.~Hencken {\it et al.},
Phys.\ Rept.\  {\bf 458} (2008) 1.

\bibitem{d'Enterria:2006ep}
D.~G.~d'Enterria,
arXiv:nucl-ex/0601001.

\bibitem{Afanasiev:2009hy}
S.~Afanasiev {\it et al.}  [PHENIX Collaboration],
Phys.\ Lett.\  B {\bf 679} (2009) 321.

\bibitem{Klein:1999qj}
S.~Klein and J.~Nystrand,
Phys.\ Rev.\  C {\bf 60} (1999) 014903; 
A.~J.~Baltz, S.~R.~Klein and J.~Nystrand,
Phys.\ Rev.\ Lett.\  {\bf 89} (2002) 012301.

\bibitem{Strikman:2005ze}
M.~Strikman, M.~Tverskoy and M.~Zhalov,
Phys.\ Lett.\  B {\bf 626} (2005) 72

\bibitem{Ivanov:2007ms}
Yu.~P.~Ivanov, B.~Z.~Kopeliovich and I.~Schmidt,
arXiv:0706.1532 [hep-ph].

\bibitem{Goncalves:2005sn}
V.~P.~Goncalves and M.~V.~T.~Machado,
J.\ Phys.\ G {\bf 32} (2006) 295;
V.~P.~Goncalves and M.~V.~T.~Machado,
arXiv:0706.2810 [hep-ph].

\bibitem{AyalaFilho:2008zr}
A.~L.~Ayala Filho, V.~P.~Goncalves and M.~T.~Griep,
Phys.\ Rev.\  C {\bf 78} (2008) 044904.

\bibitem{Baltz:2009jk}
A.~J.~Baltz, Y.~Gorbunov, S.~R.~Klein and J.~Nystrand,
Phys.\ Rev.\  C {\bf 80} (2009) 044902. 

\bibitem{Baltz:2007gs}
A.~J.~Baltz,
Phys.\ Rev.\ Lett.\  {\bf 100} (2008) 062302; 
Phys.\ Rev.\  C {\bf 80} (2009) 034901.

\bibitem{Jentschura:2009mb}
U.~D.~Jentschura and V.~G.~Serbo,
arXiv:0908.3853 [hep-ph], to appear in Eur. Phys. J C.

\bibitem{Adler:2002sc}
C.~Adler {\it et al.}  [STAR Collaboration],
Phys.\ Rev.\ Lett.\  {\bf 89} (2002) 272302.

\bibitem{Abelev:2007nb}
B.~I.~Abelev {\it et al.}  [STAR Collaboration],
Phys.\ Rev.\  C {\bf 77} (2008) 034910.

\bibitem{Klein:1999gv}
S.~R.~Klein and J.~Nystrand,
Phys.\ Rev.\ Lett.\  {\bf 84} (2000) 2330.

\bibitem{Abelev:2008ew}
B.~I.~Abelev {\it et al.}  [STAR Collaboration],
Phys.\ Rev.\ Lett.\  {\bf 102} (2009) 112301.

\bibitem{Aaltonen:2009kg}
T.~Aaltonen {\it et al.}  [CDF Collaboration],
Phys.\ Rev.\ Lett.\  {\bf 102} (2009) 242001 
[arXiv:0902.1271 [hep-ex]].

\bibitem{Aaltonen:2009cj}
T.~Aaltonen {\it et al.}  [CDF Collaboration],
Phys.\ Rev.\ Lett.\  {\bf 102} (2009) 222002.


\bibitem{Chekanov:2002xi}
S.~Chekanov {\it et al.}  [ZEUS Collaboration],
Eur.\ Phys.\ J.\  C {\bf 24} (2002) 345.

\bibitem{Aktas:2005xu}
A.~Aktas {\it et al.}  [H1 Collaboration],
Eur.\ Phys.\ J.\  C {\bf 46} (2006) 585.

\bibitem{Drees:1988pp}
M.~Drees and D.~Zeppenfeld,
Phys.\ Rev.\  D {\bf 39} (1989) 2536.

\bibitem{Cahn:1990jk}
R.~N.~Cahn and J.~D.~Jackson,
Phys.\ Rev.\ D {\bf 42} (1990) 3690; 
G.~Baur and L.~G.~Ferreira Filho,
Nucl.\ Phys.\ A {\bf 518} (1990) 786.

\bibitem{Frankfurt:2006jp}
L.~Frankfurt, C.~E.~Hyde, M.~Strikman and C.~Weiss,
Phys.\ Rev.\  D {\bf 75} (2007) 054009.

\bibitem{Klein:2003vd}
S.~R.~Klein and J.~Nystrand,
Phys.\ Rev.\ Lett.\  {\bf 92} (2004) 142003.

\bibitem{D'Enterria:2007xr}
D.~G.~.~d'Enterria {\it et al.}  [CMS Collaboration],
J.\ Phys.\ G {\bf 34} (2007) 2307.

\bibitem{Schul:2008sr}
N.~Schul and K.~Piotrzkowski,
Nucl.\ Phys.\ Proc.\ Suppl.\  {\bf 179-180} (2008) 289.

\bibitem{Alessandro:2006yt}
B.~Alessandro {\it et al.}  [ALICE Collaboration],
J.\ Phys.\ G {\bf 32} (2006) 1295; 
J.~Nystrand  [ALICE Collaboration],
Nucl.\ Phys.\ Proc.\ Suppl.\  {\bf 179-180} (2008) 156.



\end{thebibliography}
%

\end{footnotesize}


\end{document}